\documentstyle[sprocl]{article}

\input{psfig}
\def\Journal#1#2#3#4{{#1} {\bf #2}, #3 (#4)}

\def\APJ{\em ApJ}
\def\MN{\em MNRAS}

\def\be{\begin{equation}}
\def\ee{\end{equation}}
\def\bea{\begin{eqnarray}}
\def\eea{\end{eqnarray}}

\def\la{\mathrel{\mathchoice {\vcenter{\offinterlineskip\halign{\hfil
$\displaystyle##$\hfil\cr<\cr\sim\cr}}}
{\vcenter{\offinterlineskip\halign{\hfil$\textstyle##$\hfil\cr
<\cr\sim\cr}}}
{\vcenter{\offinterlineskip\halign{\hfil$\scriptstyle##$\hfil\cr
<\cr\sim\cr}}}
{\vcenter{\offinterlineskip\halign{\hfil$\scriptscriptstyle##$\hfil\cr
<\cr\sim\cr}}}}}
\def\ga{\mathrel{\mathchoice {\vcenter{\offinterlineskip\halign{\hfil
$\displaystyle##$\hfil\cr>\cr\sim\cr}}}
{\vcenter{\offinterlineskip\halign{\hfil$\textstyle##$\hfil\cr
>\cr\sim\cr}}}
{\vcenter{\offinterlineskip\halign{\hfil$\scriptstyle##$\hfil\cr
>\cr\sim\cr}}}
{\vcenter{\offinterlineskip\halign{\hfil$\scriptscriptstyle##$\hfil\cr
>\cr\sim\cr}}}}}

\begin{document}

\title{TIME--DEPENDENT HYPERCRITICAL ACCRETION ONTO BLACK HOLES}

\author{L. Zampieri
\footnote{Present address: Department of Physics, Loomis
Laboratory of Physics, University of Illinois at Urbana-Champaign,
1110 West Green Street, Urbana, Illinois, 61801-3080, U.S.A.}
}

\address{S.I.S.S.A., Via Beirut 2--4, 34013 Trieste, Italy}


\maketitle\abstracts{
Results are presented from a time--dependent, numerical investigation of
super-Eddington spherical accretion onto black holes with different
initial conditions. We have studied the stability of stationary solutions, the
non--linear evolution of shocked models and the time--dependent accretion
from an expanding medium. 
}

\section{Introduction}

Stationary spherical accretion onto black holes is certainly a
well--known and extensively studied topic (for a comprehensive review see
Nobili, Turolla and Zampieri~\cite{ntz},
Zampieri, Miller and Turolla~\cite{zmt}). 
Up to now all time--dependent investigations of spherical accretion
have been concerned either with the analysis of the stability of
isothermal~\cite{sb1}, isentropic~\cite{mo} and optically
thick\cite{gw}$^{,}$~\cite{vi2} flows or with the definition of the
parameter space within which high temperature solutions might
exist~\cite{cos}$^{,}$~\cite{st}$^{,}$~\cite{kl}. Very recently, Colpi, Shapiro
and Wasserman~\cite{csw} have studied time--dependent hypercritical
accretion of an initially expanding medium to describe the late
fall--back of material onto the compact remnant after
supernova explosion.

Despite the fact that spherical accretion has been extensively
investigated, mainly for shedding light on the efficiency of the
radiation generation, several aspects require further consideration
such as, for instance, investigating the stability properties of
high temperature solutions and searching for the existence and
non--linear evolution of possible heated or shocked models.
The results of this calculation will be presented in Sec.~2.
This study will serve also as a starting point for the investigation
of very interesting astrophysical problems, such as the late
fall--back of material onto a supernova remnant. A preliminary analysis
of this problem will be presented in Sec.~3.

\section{Stability of stationary solutions and evolution of shocked models}

We have studied the evolution of a self--gravitating, perfect hydrogen
gas in the gravitational field of a non--magnetized, non--rotating
black hole within the framework of time--dependent radiation
hydrodynamics. The present investigation has been carried out by
means of a general relativistic treatment of the radiative transfer
equation, which exploits the expansion of the specific intensity of
the radiation field into moments~\cite{th}.
A detailed review of the derivation
of the fundamental equations is presented in Zampieri, Miller and
Turolla~\cite{zmt}.
The complete
system of time--dependent, radiation hydrodynamics equations plus the
Einstein Field Equations has been solved using an original numerical
method, based on a Lagrangian finite difference scheme~\cite{zmt}.

Depending on the physical conditions
in the accretion flow, stationary solutions of
hypercritical spherical accretion onto black holes show very different
behaviours~\cite{ntz}.
In the LL (Low Luminosity) models, the inner region of the accretion flow
is in LTE with radiation at a low temperature ($T \simeq 10^5$--$10^6 K$) and
luminosity and efficiency turn out to be very small ($L
\la 4 \times 10^{33}$ erg s$^{-1}$, $e \la 10^{-7}$).
The HL (High Luminosity) models
are characterized by very high inner temperatures ($T \sim 10^9$--$10^{10}$
K). In the intermediate region ($10^3$--$10^5 r_g$, where $r_g$
is the gravitational radius), Compton cooling and free--free emission 
nearly balance Compton heating, producing a total luminosity 2--3 orders of
magnitude larger than that emitted by LL models with the
same density.

\begin{figure}
\vbox{\hskip 0.5 truecm
\psfig{figure=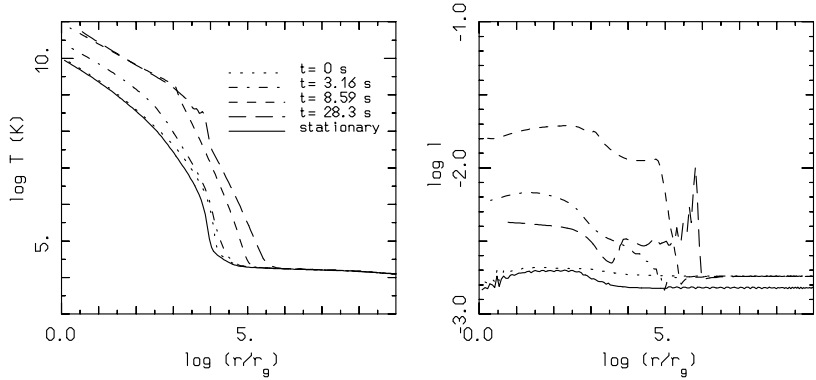,height=35truemm}}
\caption{
The gas temperature $T$ and the radiative luminosity $l$ (in Eddington units)
are plotted versus $\log (r/r_g)$ ($r_g$ is the gravitational radius)
at different times for a HL model with initial $\dot m = 28.3 $ (in Eddington
units).
}
\end{figure}

The time--dependent analysis of LL models shows that they are stable
to both thermal and radiative perturbations~\cite{zmt}.
Much more spectacular is the
phenomenology shown by HL models~\cite{zmt}.
At high accretion rates ($\dot m \ga 10$, $\dot m$ in Eddington units),
a thermal instability appears around $10^3 r_g$
after about 2--3 s, as can be seen from Fig.~1, and the
temperature increases by almost an order of magnitude.
This instability is due to the fact that the radiative heating
rate is greater than the radiative cooling rate and, at the same
time, the heating time is shorter than the dynamical time.
A few seconds after this,
the velocity profile starts to deviate significantly from free--fall
owing to the large drag exerted by the internal pressure gradients.
A compression wave develops, which becomes progressively steeper
as it propagates outward and, after 8--10 s, a hydrodynamic
shock forms.
The luminosity profile has the typical behaviour shown in Fig.~1
with the long--dashed line, showing a significant initial transient increase
lasting $\sim$ 8 s.
The shock front moves outward at an approximate speed of $10^8$ cm/s
$\simeq 10^{-2} c$. Hence this solution is definitely non--stationary.

HL models with low accretion rates ($\dot m < 10$) do not show any thermal
instability on comparable evolutionary timescales. Having lower densities, the
radiative heating and cooling are comparatively less efficient than
compressional heating and the gas is essentially adiabatic up to large radii.

\section{Spherical accretion from a uniformly expanding medium}

After (Type II) supernova explosion of a high mass star,
a reverse shock may arise from the
collision of the outflowing gas with a sufficiently massive outer
stellar envelope, leading to the formation of a dense, expanding flow
sorrounding the compact remnant.
Models for SN 1987A show that the core density is
roughly uniform ($\rho \sim 10^{-3}$ g cm$^{-3}$) and that the
initial velocity $v \propto r$ with a maximum value $\sim$ 2000 Km s$^{-1}$.
After the expansion phase, gravitationally bound shells can accrete
onto the compact remnant at hypercritical rates.
Colpi, Shapiro and Wasserman~\cite{csw} found that 
the total accreted mass is $\sim 0.1 M_\odot$ and 
could suffice to drive a remnant borne as a neutron star to further 
collapse to a black hole for a relatively soft equation of state.
In this respect we note that, for the large accretion rates occurring 
during the early phase, accretion onto a neutron star might not be 
drastically different from accretion onto a black hole because of the 
trapping of radiation and the effectiveness of neutrino losses close to 
the neutron star surface (see Chevalier~\cite{ch}).

\begin{figure}
\vbox{\hskip 0.5 truecm
\psfig{figure=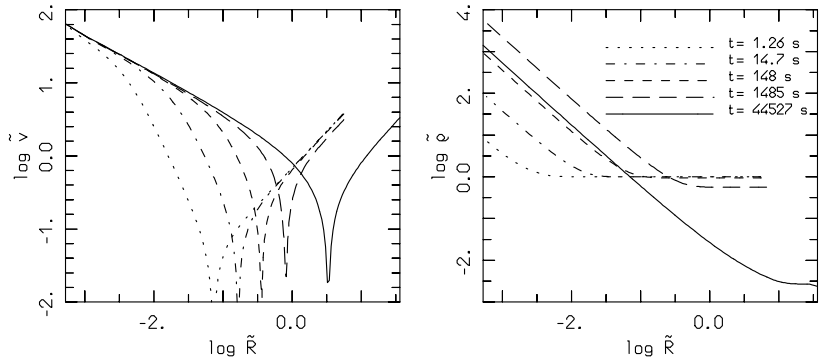,height=35truemm}}
\caption{
The gas velocity $\widetilde v$ (in units of $c_s^0$)
and the gas density $\widetilde \rho$
(in units of $\rho_0$) are plotted versus $\log {\widetilde {R}}$
(${\widetilde {R}} = r/r_a^0$, $r_a^0$ is the initial accretion radius) at 
different times. Initial 
data refer to SN 1987A: $\rho_0 \simeq 10^{-3}$ g cm$^{-3}$,
$c_s^0 = 350$ Km s$^{-1}$, $r_a^0 = 1.6 \times 10^{11}$ cm.
}
\end{figure}

Starting from the initial conditions and parameters expected for SN 
1987A~\cite{ch},
we have repeated the calculation of Colpi, Shapiro and
Wasserman for a spherically accreting 
polytropic gas with $\Gamma = 4/3$, including self--gravity.
The numerical results are shown in Fig.~2. As can
be seen looking at the velocity profile, at the beginning of the evolution
(dash--dotted line) three distinct regimes can be recognized:
free--fall in the region below the accretion radius $r_a$,
near hydrostatic equilibrium between $r_a$ and the marginally bound 
radius $r_{mb}$ (below which mass elements are gravitationally bound)
and uniform expansion above $r_{mb}$.
At $t \simeq 4700$ s, the accretion rate
reaches its maximum value, $\dot m \simeq 2 \times 10^{10}$.
The late fall back can be described in
terms of a dust infall in which only those shells within $r_{mb}$ are
eventually accreted~\cite{csw}.
The total accreted
mass is $M_{acc} \simeq 0.05 M_\odot$, 
in agreement with the value estimated by Colpi, Shapiro and Wasserman.
Then, although the initial mass contained in the expanding shells is $\sim 5 
M_\odot$, only a small fraction is accreted. Since for the Birkhoff 
theorem the mass outside $r_a$ has no effect on the motion of the inner 
gas and since $M_{acc}$ is very small in comparison with the mass of the 
central black hole, self--gravity does not play any important role.

The present simple hydrodynamic model of spherical accretion from a
uniformly expanding polytropic gas represents the starting point of
this investigation and does not include the transfer of radiation,
the role of which might probably become important. In fact, although
matter and radiation are presumably in LTE and the efficiency of accretion
onto black holes is very low, the actual value of the accretion rate
is so high that near Eddington luminosities might be produced, strongly
influencing the dynamics of the accretion flow and hence the total
accreted mass. This study along with a companion analysis of spherical
hypercritical accretion onto neutron stars are presently under way.

\section*{Acknowledgments}

I would like to thank Stu Shapiro and Monica Colpi for suggesting that I
investigate spherical accretion from a uniformly expanding medium and for
their continuing guidance. I also thank Roberto Turolla and Ira Wasserman
for useful discussions. This research was supported in part by NSF grant
AST 93-15133 and NASA grant 5-2925 at the University of Illinois.

\end{document}